# Modified one dimensional elastic wave equations that retain time synchronization under spatial coordinate transformations

Ruiwen Yao, Zhihai Xiang[*]

AML, Department of Engineering Mechanics, Tsinghua University, Beijing 100084, China

**Abstract**

In contrast to the traditional elastodynamic equations, a more comprehensive formulation of one dimensional (1D) elastodynamic equations is given for inhomogeneous media by using the coordinate transformation method. These modified equations consider the gradient of pre-stresses so that they are form-invariant and can retain time synchronization under spatial coordinate transformation, which comply with *the principle of general invariance*. A numerical example is conducted to compare the distributions of wave speeds calculated by the modified equations and the traditional equations. It demonstrates that the traditional equations are good approximations of the modified equations only when the wave frequency is sufficiently high.

---

[*] Email: xiangzhihai@tsinghua.edu.cn

## 1. Introduction

In the development of meta-devices that can steer wave propagations, coordinate transformation method plays an important role in determining the metamaterial parameters [1-6]. This method transforms the governing equations from a virtual space to a physical space of interest. If the equation forms are invariant after coordinate transformation, the material parameters in the physical space can be obtained by comparing the equations before and after transformation. Usually, the media in the virtual space are homogeneous, while the media in the physical space are inhomogeneous and are commonly called 'transformation media'. Actually, transformation methods can also be interpreted in the language of general relativity [7]. The prerequisite to use these methods is the form-invariant property of the equations under a general spatial transformation, which has been verified for electromagnetic and acoustic waves [1,5,6]. However, Milton et al. [8] demonstrate that the traditional elastodynamic equations are not form-invariant, implying the impossibility of designing an ideal elastic meta-device based on these equations. On the other hand, *the principle of general invariance* states that all laws of physics must be invariant under general coordinate transformations [9]. Therefore, the traditional elastodynamic equations may fail to describe elastic wave propagation in transformed inhomogeneous media, although they are widely recognized to be absolutely accurate in homogenous media. In this sense, the Willis equations could be good choices to design meta-devices because they are form-invariant with symmetric elastic tensor [8,10,11]. However, people are still used to using the traditional equations for the simplicity. For example, Farhat et al. [12] achieve a cylindrical cloak to control bending waves in thin elastic plates; Hu et al. [13] and Chang et al. [14] control elastic waves by introducing a locally affine mapping instead of a general transformation; Norris and Parnell [15] has explored hyperelastic cloaking by employing statically equilibrated pre-stresses.

This paper presents modified 1D elastic wave equations that are form-invariant and real-valued. The following text will illustrate that the traditional equations are special forms of these modified equations for homogeneous media. In contrast to the traditional equations, the additional term in the modified equations are related with the gradient of pre-stresses. To give a concise and straightforward comparison between these two sets of equations, only the P wave propagation is discussed in this paper.

## 2. The modified equations

In particular, the traditional equations in 1D virtual space in the absence of body forces can be written as

$$\bar{\sigma}(\bar{x},t) = \bar{C}\frac{\partial \bar{u}(\bar{x},t)}{\partial \bar{x}}, \tag{1a}$$

$$\frac{\partial \bar{\sigma}(\bar{x},t)}{\partial \bar{x}} = \bar{\rho}\frac{\partial^2 \bar{u}(\bar{x},t)}{\partial t^2}, \tag{1b}$$

where $\bar{x}$ is the virtual space coordinate; $t$ denotes time; $\bar{u}$ is the displacement along coordinate axis; $\bar{\sigma}$ is the longitudinal stress; $\bar{C}$ is the stiffness and $\bar{\rho}$ is the mass density. These equations are certainly rigorous in homogeneous media, in which the stiffness and the mass density are constants. In this paper, variables and constants except for the time in the virtual space are denoted with superposed bars. The variables without superposed bars are in the physical space. After applying a general mapping from the virtual space to the physical space: $x = f(\bar{x})$, the traditional equations are expected to change their forms.

For the general 3D case, Norris and Shuvalov [10] have pointed out that one can obtain a symmetric elasticity tensor if assuming the field variables before and after transformation satisfy $\boldsymbol{u}(\boldsymbol{x}) = \boldsymbol{F}^{-T} \cdot \bar{\boldsymbol{u}}(\bar{\boldsymbol{x}})$ and $\boldsymbol{\sigma}(\boldsymbol{x}) = \boldsymbol{F} \cdot \bar{\boldsymbol{\sigma}}(\bar{\boldsymbol{x}}) \cdot \boldsymbol{F}^T / \det(\boldsymbol{F})$ [8, 12], where $\boldsymbol{u}$ is the displacement vector; $\boldsymbol{\sigma}$ is the stress tensor; $\boldsymbol{x}$ is the position vector; and $\boldsymbol{F}$ is the deformation gradient with elements $F_{ij} = \partial x_i / \partial \bar{x}_j$. Therefore, for this 1D problem one can assume

$$u(x,t) = \bar{u}(\bar{x},t) / f'(\bar{x}), \tag{2a}$$

$$\sigma(x,t) = f'(\bar{x}) \bar{\sigma}(\bar{x},t), \tag{2b}$$

where $f'(\bar{x})$ is the derivative of $f$ over $\bar{x}$.

Substituting Eq. (2) into Eq. (1), and taking a note of the chain rule $\frac{\partial}{\partial \bar{x}}(\ ) = f'(\bar{x}) \frac{\partial}{\partial x}(\ )$, one can obtain modified 1D elastic equations in the physical space as

$$\sigma(x,t) = C(x) \frac{\partial u(x,t)}{\partial x} + S(x) u(x,t), \tag{3a}$$

$$\frac{\partial \sigma(x,t)}{\partial x} = \rho(x) \frac{\partial^2 u(x,t)}{\partial t^2} + S(x) \frac{\partial u(x,t)}{\partial x} + K(x) u(x,t), \tag{3b}$$

where

$$C(x) = \bar{C}[f'(\bar{x})]^3, \quad \rho(x) = \bar{\rho} f'(\bar{x}), \quad S(x) = \bar{C} f'(\bar{x}) f''(\bar{x}), \quad K(x) = \bar{C}[f''(\bar{x})]^2 / f'(\bar{x}). \tag{4}$$

Compared with the traditional equations, Eq. (3) has three additional terms: $Su$, $S\partial u/\partial x$ and $Ku$, which are generally nonzero in inhomogeneous media if $f''(\bar{x}) \neq 0$. If the mapping function is linear, the transformed media must be homogeneous. In this case, Eq. (3) are the same as the traditional equations, because $f'(\bar{x}) = const$ and $f''(\bar{x}) = 0$. To verify the form-invariance of the modified wave equations, one can write Eq. (3) in the virtual space by simply adding superposed bars on the variables, and then find their unchanged forms in the physical space by using Eq. (2). Moreover, the modified equations are real-valued. These features indicate that the modified equations satisfy *the principle of general invariance*, while the traditional equations are not.

The first equation in Eq. (3) is the modified constitutive relation that defines the stress increment due to an infinitesimal deformation in inhomogeneous media. It is different from the traditional constitutive relation in homogenous media with an additional term $Su$, which contains the information of pre-stress gradient $S(x)$ [16]. Actually, the pre-stress is the concomitant of inhomogeneity. The similar pre-stress effects due to the coordinate transformation are also reported in [15] and [17].

The additional terms $S\partial u/\partial x$ and $Ku$ in the equilibrium equation in Eq. (3) are the consequent results when substituting the modified constitutive equation into the equilibrium equation.

As aforementioned, the traditional equations are commonly used to approximately describe the wave propagation in inhomogeneous media. By removing the additional terms in Eq. (3), one can obtain the traditional equations for 1D elastic waves in the inhomogeneous transformation media

$$\sigma(x,t) = C(x)\frac{\partial u(x,t)}{\partial x}, \tag{5a}$$

$$\frac{\partial \sigma(x,t)}{\partial x} = \rho(x)\frac{\partial^2 u(x,t)}{\partial t^2}. \tag{5b}$$

And the expressions of $\rho(x)$ and $C(x)$ are the same as those in Eq. (4) for the transformation media. Hu et al. [13] have obtained Eq. (5) in 3D space by using the concept of local affine and considering the conservation of kinetic and strain energies. They point out that this approximation method can give a satisfactory control of elastic waves at high frequency or weak heterogeneity. The following sections will show how Eq. (3) differs from Eq. (5) in the description of 1D elastic wave propagations in inhomogeneous media. By comparing elastic wave speeds in the physical space, one can also find Eq. (5) would become more accurate at higher frequency or in weaker heterogeneity. Actually, higher frequency and weaker heterogeneity convey the same concept, because a higher frequency wave would experience less material fluctuations in its shorter wave length.

## 3. Wave speed

Eliminating $\sigma(x,t)$ in Eq. (3), obtains

$$C(x)u_{,xx}(x,t) + C_{,x}(x)u_{,x}(x,t) + R(x)u(x,t) = \rho(x)\frac{\partial^2 u(x,t)}{\partial t^2}, \tag{6}$$

where $\phi_{,x}$ denotes the derivative of function $\phi$ over $x$; and $R(x) = S_{,x}(x) - K(x)$. Actually, $R(x)u(x,t)$ can be regarded as the effective body force due to the gradient of the pre-stress [16].

According to Bloch's theorem [18], the wave amplitude $a(x)$ should be considered as a function of position for periodic media. Since the periodic medium can be regarded as a special case of inhomogeneous media, one can try to extend this assumption for general inhomogeneous media discussed in this paper. Furthermore, the initial wave phase should be calculated as an integration of wave number $p(x)$ along the distance from a reference position $x_0$ with zero phase to current position. Therefore, the displacement of monochrome vibrations in inhomogeneous media can be written as

$$u(x,t) = a(x)\exp\left\{i\left[\int_{x_0}^{x} p(s)\,ds - \omega t\right]\right\}, \tag{7}$$

where $\omega$ is the angular frequency. This equation is also correct for homogenous media, where the wave amplitude and wave number are constants.

Substituting Eq. (7) into Eq. (6) and separating real and imaginary terms, obtains

$$\left[R(x) - C(x)p^2(x) + \rho(x)\omega^2\right]a(x) + C_{,x}(x)a_{,x}(x) + C(x)a_{,xx}(x) = 0, \tag{8}$$

$$C_{,x}(x)a(x)p(x) + 2C(x)a_{,x}(x)p(x) + C(x)a(x)p_{,x}(x) = 0. \tag{9}$$

Eliminating $a_{,x}(x)$ and $a_{,xx}(x)$ from Eq. (8) and Eq. (9), obtains

$$R(x)+\rho(x)\omega^2-C(x)p^2(x)-C_{,x}(x)\frac{[C(x)p(x)]_{,x}}{2C(x)p(x)}-C(x)\frac{[C(x)p(x)]_{,xx}}{2C(x)p(x)}+3C(x)\left\{\frac{[C(x)p(x)]_{,x}}{2C(x)p(x)}\right\}^2=0. \quad (10)$$

Further substituting the fundamental relationship between the wave number and the wave speed $p(x)=\omega/v(x)$ into Eq. (10), one obtains the wave speed equation which is much more complex than that in homogeneous media

$$\left\{R(x)-\frac{1}{2}C_{,xx}(x)+\frac{1}{4}\frac{[C_{,x}(x)]^2}{C(x)}\right\}v^2(x)-\frac{C(x)}{4}[v_{,x}(x)]^2+\frac{C(x)}{2}v_{,xx}(x)v(x)+\omega^2\left[\rho(x)v^2(x)-C(x)\right]=0. \quad (11)$$

Since here considers only the pure P wave rather than the guided wave, there is no speed-frequency dispersion brought by surface and interface conditions. In addition, since only the pure linear elasticity rather than the viscoelasticity is considered, there is no speed-frequency dispersion brought by the time-dependent properties of the media. Thus, the wave speed should be independent with the frequency and both of the following conditions should be satisfied

$$\rho(x)v^2(x)-C(x)=0, \quad (12)$$

$$\left\{R(x)-\frac{1}{2}C_{,xx}(x)+\frac{1}{4}\frac{[C_{,x}(x)]^2}{C(x)}\right\}v^2(x)-\frac{C(x)}{4}[v_{,x}(x)]^2+\frac{C(x)}{2}v_{,xx}(x)v(x)=0. \quad (13)$$

Eq. (12) clearly shows

$$v(x)=\sqrt{\frac{C(x)}{\rho(x)}}, \quad (14)$$

which is similar to the wave speed in homogeneous media. For inhomogeneous transformation media, Eq. (13) can be proven to be an identity equation by substituting Eqs. (4) and (14) into it.

## 4. Time synchronization

The concise form of Eq. (14) has profound significance because it keeps time synchronization between the virtual space and the physical space. Substituting Eq. (4) into Eq. (14), obtains

$$v(x)=f'(\bar{x})\sqrt{\frac{\bar{C}}{\bar{\rho}}}=f'(\bar{x})\bar{v}, \quad (15)$$

where $\bar{v}$ is the wave speed in the virtual space. When an interval $[\bar{x}_1,\bar{x}_2]$ in the virtual space is mapped to the interval $[x_1,x_2]$ in the physical space, the elastic wave must spend the same time to travel through them, because

$$\int_{x_1}^{x_2}\frac{\mathrm{d}x}{v(x)}=\int_{\bar{x}_1}^{\bar{x}_2}\frac{f'(\bar{x})\mathrm{d}\bar{x}}{v(x)}=\int_{\bar{x}_1}^{\bar{x}_2}\frac{\mathrm{d}\bar{x}}{\bar{v}}=\frac{\bar{x}_2-\bar{x}_1}{\bar{v}}. \quad (16)$$

Because only spatial transformation is used, time synchronization should be preserved before and after the transformation. Furthermore, since $p(x)=\omega/v(x)$ and $\bar{p}=\omega/\bar{v}$, the phase synchronization can be easily verified as $\int_{x_1}^{x_2}p(x)\mathrm{d}x=(\bar{x}_2-\bar{x}_1)\bar{p}$ by using Eq. (16). This coincides with Eq. (7). Further considering Eq. (2a), the wave amplitude should satisfy $a(x)=\bar{a}/f'(\bar{x})$, where $\bar{a}$ denotes the amplitude in the virtual space.

## 5. Numerical comparison between the traditional and the modified equations

In contrast, the traditional equations do not generally keep the time synchronization, because Eq. (14) cannot be obtained from Eq. (5). Actually, the Lamé-Navier equation corresponding to Eq. (5) is

$$C(x)u_{,xx}(x,t) + C_{,x}(x)u_{,x}(x,t) = \rho(x)\frac{\partial^2 u(x,t)}{\partial t^2},$$

which is different from Eq. (6) for the lack of $R(x)u(x,t)$. After the similar procedure from Eq. (6) through Eq. (11), one can obtain the wave speed equation as

$$\left\{-\frac{1}{2}C_{,xx}(x) + \frac{1}{4}\frac{[C_{,x}(x)]^2}{C(x)}\right\}v^2(x) - \frac{C(x)}{4}[v_{,x}(x)]^2 + \frac{C(x)}{2}v_{,xx}(x)v(x) + \omega^2\left[\rho(x)v^2(x) - C(x)\right] = 0. \qquad (17)$$

It turns out that the above equation is a simplified version of Eq. (11) by removing the adjustment term $R(x)$. Because

$$\left\{-\frac{1}{2}C_{,xx}(x) + \frac{1}{4}\frac{[C_{,x}(x)]^2}{C(x)}\right\}v^2(x) - \frac{C(x)}{4}[v_{,x}(x)]^2 + \frac{C(x)}{2}v_{,xx}(x)v(x) \neq 0 \;,\quad \text{Eq. (17) definitely leads to}$$

speed-frequency dependency. Actually, referring Eq. (4), one can obtain $R(x) = \bar{C}f'''(\bar{x})$. This means the wave speed equations of the modified equations and the traditional equations are inherently different, unless $f'''(\bar{x}) = 0$.

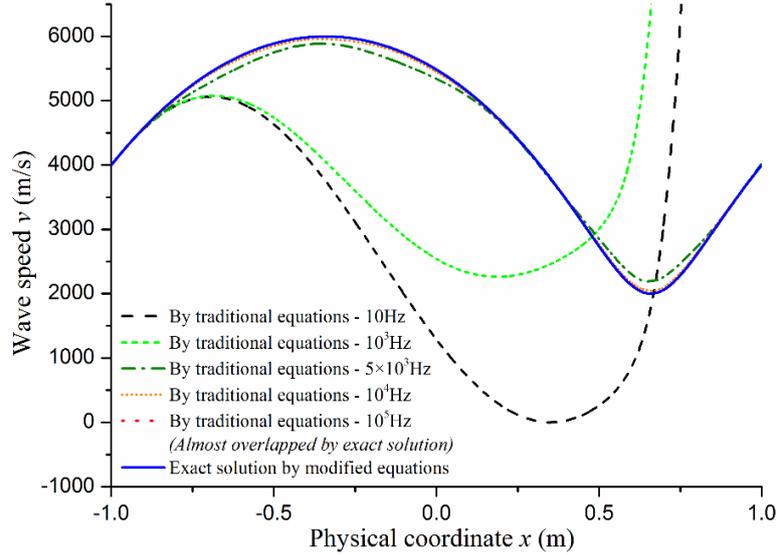

**Fig. 1.** Comparison between the wave speed distributions determined by the traditional equations and the modified equations under different frequencies.

It notices that the last term of Eq. (17) is dominant at very high frequency. Thus Eq. (14) can be approximately satisfied and the traditional equations are good approximations of the modified equations. This can be demonstrated in the following example, where the mapping function is $x = f(\bar{x}) = \bar{x} + \frac{1}{2\pi}\left[\cos(\pi\bar{x}) + 1\right]$, $(x, \bar{x} \in [-1,1])$; the wave speed in the virtual space is $\bar{v} = \sqrt{\bar{C}/\bar{\rho}} = 4000$ (m/s); and the boundary conditions are

$$v(-1) = \left[ f'(\bar{x}) \sqrt{\frac{\bar{C}}{\bar{\rho}}} \right]_{x=-1} = \sqrt{\frac{\bar{C}}{\bar{\rho}}} = 4000 \text{ (m/s)} \quad \text{and} \quad v_{,x}(-1) = \left[ \frac{f''(\bar{x})}{f'(\bar{x})} \sqrt{\frac{\bar{C}}{\bar{\rho}}} \right]_{x=-1} = \frac{\pi}{2} \sqrt{\frac{\bar{C}}{\bar{\rho}}} = 2000\pi \text{ (s}^{-1}\text{)}.$$

By using the fourth-order Runge-Kutta method, one can numerically solve the wave speeds determined by Eq. (11) and Eq. (17), respectively. As a result, the solution from Eq. (11) always coincides with Eq. (15) and does not vary over frequencies (see the blue solid line in Fig. 1). In contrast, the wave speed from the traditional equations does vary over frequencies, especially at low frequencies. The lower the frequency, the larger the deviation from the exact solution of Eq. (15). When the frequency is too low, the speed curve even behaves some instability. However, if the frequency is sufficiently high, the deviation can be very small or even negligible, which is also consistent with the report of Hu et al [13].

## 6. Concluding remarks

In summary, this paper gives modified 1D elastic wave equations by coordinate transformation method, which can precisely keep time synchronization between the virtual space and the physical space. The numerical example shows that the traditional equations could give good approximations for inhomogeneous media only when the wave frequency is sufficiently high. In contrast, the modified equations can give accurate results independent of frequencies. These findings could be important to the applications using low frequency elastic waves, such as seismic exploration.

By considering the additional terms in the modified equations, people could make a better prediction of the behaviours of seismic waves in geophysics. However, as pointed out by [16], these additional terms are related with the gradient of the pre-stress, which could be difficult to be determined in practice. For example, the crust of earth contains very complicated and randomly distributed inhomogeneous rocks, soils, liquids, gases, minerals and faults, etc. In this scenario, some inverse analyses based on the measured waveform could give a help to identify these additional terms. All in all, the main purpose of this paper is to draw people's attention to the accuracy of elastic wave equation for inhomogeneous media.

**Acknowledgements**

The authors are grateful for the support from the National Science Foundation of China through grant 11272168.